\documentclass[aps,twocolumn,superscriptaddress]{revtex4-1}
\usepackage{amsmath,amssymb}
\usepackage{graphics,graphicx}
\usepackage{dcolumn,bm}
\usepackage{psfrag}
\usepackage{color}
\usepackage{multirow}
\newcommand{\cu}
{\affiliation{Department of Physics, University of Calcutta,
92 Acharya Prafulla Chandra Road, Kolkata 700009, India.}}

\newcommand{\imsc}
{\affiliation{The Institute of Mathematical Sciences, CIT Campus, Taramani, Chennai 600113, India.}}

\newcommand{\hbni}
{\affiliation{Homi Bhabha National Institute, Training School Complex, Anushakti Nagar, Mumbai 400094, India.}}

\newcommand{\epsi}
{\epsilon}
\begin{document}
\title{Effect of presence of rigid impurities in a system of annihilating domain walls with dynamic bias}

\author{Reshmi Roy}
\imsc
\hbni
\email{reshmiroy@imsc.res.in}
\author{Parongama Sen}%
\cu

\begin{abstract}
The dynamics of interacting domain walls, regarded as a system of particles which are biased to move towards their nearest  neighbours and annihilate when they meet, have been studied 
 in the recent past.
 We study the effect of the presence of 
 a fraction $r$ of quenched impurities (which act as rigid walkers) on the dynamics. Here, in case  two domain walls or one impurity and one domain wall happen to be on the same site, both get simultaneously annihilated. It is found that for any non-zero value of $r$, the dynamical behaviour changes as the
 surviving fraction of particles $\rho(t)$  attains a constant value. $\rho(t)t^\alpha $ shows a universal behaviour when plotted against $r^\beta t$
 with $\alpha, \beta$ values depending on whether the particles are rigid or nonrigid. Also,
 the values differ for the biased and unbiased cases. The time scale associated with the particle decay obtained in several ways shows that it varies with $r$ in a power law manner with a 
universal exponent.
\end{abstract}

\maketitle

\section{Introduction}
 Domain wall dynamics  in magnetic and other systems is an important area of research.
Such dynamics may include domain wall nucleation, propagation, annihilation etc. induced by some  technological processes. 
In simple magnetic systems described by Ising spins, the interfaces separate the domains of up and down spins and at zero temperature, ideally no such interface can exist. However,
preparing the system at a high temperature where such interfaces are numerously present and subsequently quenching it to zero temperature, domain growth phenomena can be observed. 
In a one dimensional Ising model governed by Glauber dynamics, the evolution  of the system can be  equivalently represented by 
the dynamics of diffusing particles which annihilate when two such particles  meet \cite{privman, liggett, krapivsky,odor}. 
In infinite systems, the domains may  continue to grow indefinitely, such that the system is always in an active state. 
Presence of impurities or defects, non-ferromagnetic in nature at a point, can impede the domain 
growth and one can regard the system having a domain wall pinned at that point \cite{volkov,karna,ghosh,Blachowicz,sbiswas,moglia}. This may lead to an absorbing state. However, it is possible to achieve a depinning transition by appropriate means in real systems which again drives the system to an active state. 


In order to mimic the presence of such defects occurring between two domains, we consider 
a system of diffusing annihilating particles representing the domain walls as in the Ising-Glauber model,  where a finite fraction of such particles  is immobile or rigid. A rigid particle, representing a defect, can, however, vanish due to annihilation as it meets a non-rigid one. Note that no external 
agent is required for the domain wall motion or annihilation in the kinetic Ising model. 
The above scenario can be 
regarded as a reaction diffusion system of one species of the form $A+ A \to \emptyset$ which has been studied in different contexts in the past \cite{privman, liggett, krapivsky,odor}. 


In some previous works, 
a  $A+A \to \emptyset$ system in the presence of a dynamic bias where  the particles have a bias
to move towards their nearest neighbour has been studied in detail \cite{sb_ps_pray,ps_pray,daga,rr_pray_ps,rr_ps,park1,park2,rr_ps_pray}. In the present work, 
we include this bias and consider the effect of the presence of the  rigid impurities/walls/particles in the sense  they never change
their position with time.  The aim is to 
see how the system is affected by the presence of the rigid walls, in particular, whether there is any active-absorbing phase transition. 

The main point of interest is how the particle density decays with time and as mentioned earlier, for
an active state, it typically shows an algebraic decay while for an absorbing state it is 
expected to reach a non-zero constant value. 
In addition to domain growth phenomena, one can also study 
another important dynamical feature going by the name persistence. 
In spin models, this is  estimated by the fraction of spins which never change sign till a particular time. 
In the equivalent particle picture, this is estimated by the number of sites on the lattice 
never visited till that time. The relevant  exponents of the Ising Glauber model for the unbiased case are known exactly and for the biased case these have been estimated by numerical methods.

\section{Model and quantities calculated} \label{model}

We consider a lattice of size $L$ occupied by 
$\rho L$ particles initially. Let the fraction of the rigid particles be denoted by $r$. Though the rigid walkers
do not move, they can be annihilated by a  non-rigid particle when the latter moves to the same site occupied by the former, so that both the densities are functions of time. 
The non-rigid particles move towards their nearest
neighbour with probability $0.5+\epsilon$ and with the complementary probability it moves in the opposite direction.
The motion is purely diffusive for $\epsilon=0$ while for $\epsilon=0.5$
it is fully deterministic. 

In absence of the rigid particles, the density of surviving particles $\rho(t)$ shows a power law decay with time. For 
$\epsilon=0$,  $\rho(t) \sim t^{-\frac{1}{2}}$ is an exact result \cite{krapivsky}. This power law exponent changes to $1$ for any $\epsilon >  0$ 
\cite{sb_ps_pray,ps_pray}. The persistence probability  also shows a power law decay with time 
with different exponents for $\epsilon=0$ and $\epsi \neq 0$. Other quantities studied include the probability distribution of the displacements, probability of
changing direction etc.  Additional variations, like having repulsive (instead of attractive) dynamics using a negative value of  $\epsilon$, different schemes of updating, were considered as well \cite{rr_pray_ps,rr_ps, rr_ps_pray}. It was found by tagging the particles that the walk is ballistic in nature for the attractive bias while the displacements follow a double peaked non-Gaussian behaviour \cite{rr_pray_ps}. In this article, we are interested to find how the behaviour of the relevant quantities change in presence of the 
rigid walkers. Obviously, if the rigid fraction is small, it may happen that the rigid particles are annihilated fast enough such that at long times the behavior of $r=0$ will prevail. 
On the other hand, for a large fraction $r$ it is expected that at long times only some rigid agents will remain in the system.

The positions of the particles are updated asynchronously and
the dynamics are as follows: in a lattice of size $L$, a site is chosen randomly; if there is a non rigid particle it hops one step in the direction of its nearest neighbour with probability $0.5+\epsilon$ and with the complementary probability it moves in the opposite direction. For two equidistant nearest neighbours, it moves to either direction with equal probability $\frac{1}{2}$. 
If the destination site is occupied by any particle, either rigid or non-rigid, then both  the
particles are annihilated simultaneously. While updating the position of the particles, if  there is a rigid particle or no particle in the selected site, no movement is made. $L$ number of updates constitute one Monte Carlo (MC) step. Here, $\epsilon$ is considered to be non-negative $(0 \leq \epsilon \leq 0.5)$ and for $\epsilon=0$, particles perform purely random walk, whereas the particles move deterministically towards their nearest neighbour for $\epsilon=0.5$ as mentioned earlier.

We are primarily interested to estimate the time dependence of the fraction of the surviving particles, denoted by $\rho(t)$, one of the most important issues in such systems, that
determines the universality.
After completing each MC step, we have calculated the number of surviving particles at that time
scaled by the total number of particles present initially. We have also studied the decay of the rigid and non rigid particles termed $\rho_r(t)$ and $\rho_{nr}(t)$ separately, the scaling factors being the number of the respective types of particles at $t=0$. Next, the persistence probability $P(t)$ is studied which  in the lattice walker system is defined as the probability
that a site has not been visited by any of the particles till time $t>0$. Initially, at $t=0$, all sites are regarded as unvisited. 

The simulations have been conducted on a maximum system size $L=20000$ randomly half filled with $10000$ walkers initially and the number of realisations was 1000. Periodic boundary conditions have been imposed in all the cases.

\section{Results}

\subsection{Dynamics of particle decay}

For very small values of the rigidity $r$,
 $\rho(t)$ shows a power law decay; $\rho(t) \sim t^{-0.5}$ for $\epsilon=0$ and $\rho(t) \sim t^{-1}$ when $\epsilon>0$. This implies
 that $\rho(t)$ retains its usual power law behaviour for small values of $r$ $(\mathcal O(10^{-2}))$. As $r$ increases, $\rho(t)$ tends
 to attain a saturation value $\rho_{sat}$ after a slow decay in time. With the increase in $r$, the saturation is reached faster and 
the saturation value is also larger. This is true for any $\epsilon \geq 0$. However, we will agrue later that for any $r > 0$, there will be a deviation from the $r=0$ behavior.

{\textit {Notations and general behavior}}: Denoting  any of the densities (total, rigid or non-rigid particle) by a generic symbol $\rho_g$, we observe that a data collapse can be obtained by plotting $\rho_gt^{\alpha}$  against $r^{\beta}t^\alpha$. Therefore, a general
scaling form can be written as

\begin{eqnarray}
\rho_g(t) \sim t^{-\alpha}f(r^\beta t^\alpha).
\label{rho_fit_generic}
\end{eqnarray} 
When the density shows a saturation value, say $\rho_{gs}$, then, for a particular value of $r$, we obtain the following functional dependence on $t$
\begin{eqnarray}
\rho_g(t) - \rho_{gs} \sim  t^{-a} \exp(-t^b/\zeta).
\label{rhosat_fit_generic}
\end{eqnarray} 

The compatibility of the above two equations necessitates that $\alpha = a$ and also indicates that there is a timescale varying as $r^{-\beta/\alpha}$ such that $\zeta^{1/b}$, which represents a timescale in equation \ref{rhosat_fit_generic}, must vary in the same way. We will report results which are consistent with this. 

In the next two subsections, we discuss the results for $\epsilon=0$ and 
$\epsilon \neq 0$ separately. 

\subsubsection{$\epsilon=0$}

\begin{figure}[!ht]
\includegraphics[width=7cm]{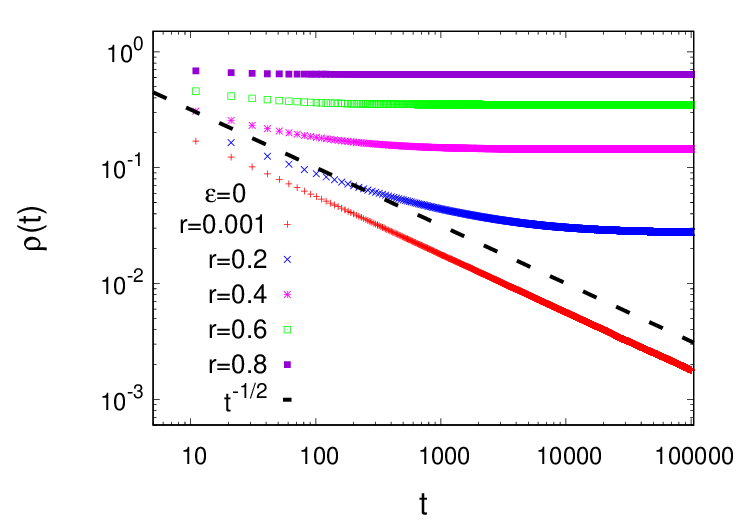} 
\caption{$\rho(t)$ vs $t$ for $\epsilon=0$}
\label{rho_ep0}
\end{figure}

\begin{figure}[h]
\includegraphics[width=7cm]{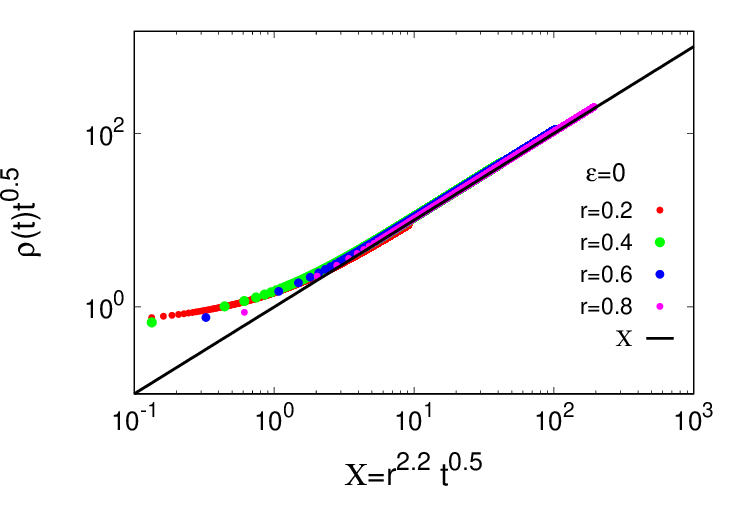}
\caption{Data collapse of $\rho(t)t^\frac{1}{2}$ vs $r^{2.2}t^\frac{1}{2}$ for $\epsilon=0$.}
\label{alive_collapse_ep0}
\end{figure}

Fig. \ref{rho_ep0} shows
$\rho(t)$ for $\epsilon=0$ for several values of rigidity $r$. One can observe that apparently, for very small $r \sim {\mathcal O}(10^{-2})$, the behavior is still a power law with the exponent value close to 1/2 while for larger values of $r$, the expected saturation in time is shown. A data collapse of $\rho(t)$, shown in Fig. \ref{alive_collapse_ep0}, can be obtained using the scaling form of Eq. \ref{rho_fit_generic} where $\alpha \simeq 0.5$ and $\beta \simeq 2.2$ such that one can argue that 
\begin{eqnarray}
 \rho(t) \sim t^{-0.5}f(r^{2.2} t^{0.5}).
\label{rho_fit_ep0}   
\end{eqnarray}
We note that  till $r^{2.2} t^{0.5} \approx 1$, the scaling function is nearly a constant,
which indicates the behaviour of $\rho(t)$ will closely follow $\rho(t) \sim t^{-1/2}$ which is 
true for $r=0$. Beyond this value, the scaling function is linear such that $\rho(t)$ is a constant. This crossover occurs at $t \approx \frac{1}{r^{4.4}}$ which is very large for small $r$ values and therefore cannot be detected in the simulations. Our results in this subsection 
would henceforth be presented for $r \geq 0.2$ only.  

\begin{figure}
    \centering
     \includegraphics[width=7cm]{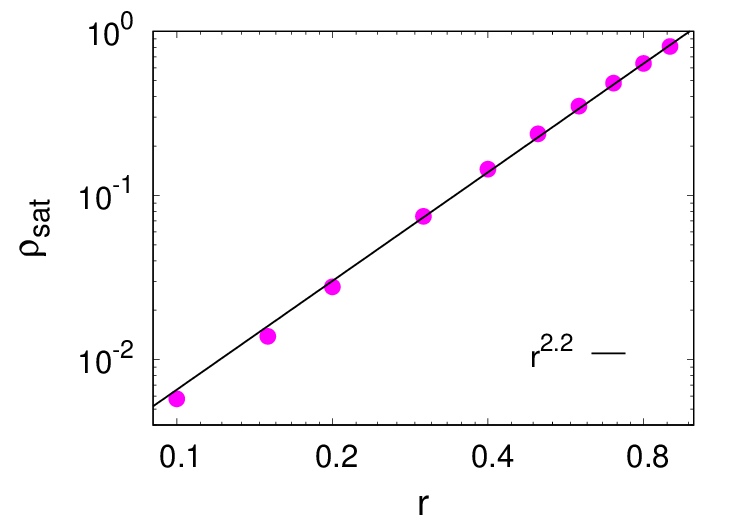}
    \caption{ $\rho_{sat}$ vs $r$ for $\epsilon = 0$ showing power law variation as mentioned
in the key.}
    \label{rhosat_ep0}
\end{figure}

Since at large times $\rho(t)$ attains a constant value $\rho_{sat}$, $f(z)$ should follow the behavior $f(z) \sim z$ at large times 
where $z=r^\beta t^\alpha$. This indicates  $\rho_{sat}$ should be
proportional to  $r^{\beta} = r^{2.2}$.
 The 
variation of the saturation value $\rho_{sat}$ shows that it  increases with $r$, and indeed shows a power law behaviour with an exponent close to 2.2 (see Fig. \ref{rhosat_ep0}). 

Keeping $r$ fixed, when this saturation value $\rho_{sat}$ is subtracted from $\rho(t)$, the data can be fit to the form of Eq. \ref{rhosat_fit_generic} with $a=0.5$, such that 
\begin{equation}
\rho(t) - \rho_{sat} = k\ t^{-0.5} \exp(-t^b/\zeta),
\label{rhosat_eq_ep0}
\end{equation} 
where $k$ is the proportionality constant. 
As noted already, $\tau \propto \zeta^{\frac{1}{b}}$ can be identified as a timescale.

A similar study of the densities for the  rigid 
and non-rigid particles has been made separately. 
In case there is a steady state, there can be no non-rigid particles left in the system  and the density of the surviving rigid particles will 
depend  on the value of $r$. Hence one can expect a different dynamical behaviour for the two types of particles.

$\rho_{nr}(t)$, the density of non-rigid particles at time $t$, is  estimated by scaling  the number of surviving non-rigid particles by the initial number of
such particles in the system. The decay of $\rho_{nr}(t)$ can be fitted to Eq. \ref{rhosat_fit_generic} where $\rho_g(r)=\rho_{nr}(t)$ and $\rho_{gs} =0$ in this case. Hence  the variation with time for a fixed $r$ value can be written as
\begin{eqnarray}
  \rho_{nr}(t)=k_{nr} t^{-0.5}\exp(-t^{b_{nr}}/\zeta_{nr}),
\label{nonrigid_eq_ep0}
\end{eqnarray}  
with $k_{nr}$
a constant.
From the above variation,  another timescale $\tau_{nr} \propto \zeta_{nr}^{1/b_{nr}}$, can be obtained.

\begin{figure}[h]
\includegraphics[width=4.2cm]{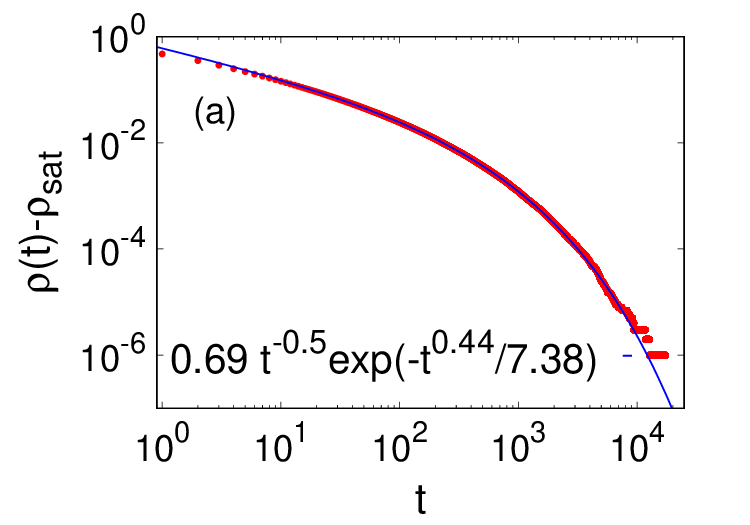} \hspace{-0.4cm}
\includegraphics[width=4.2cm]{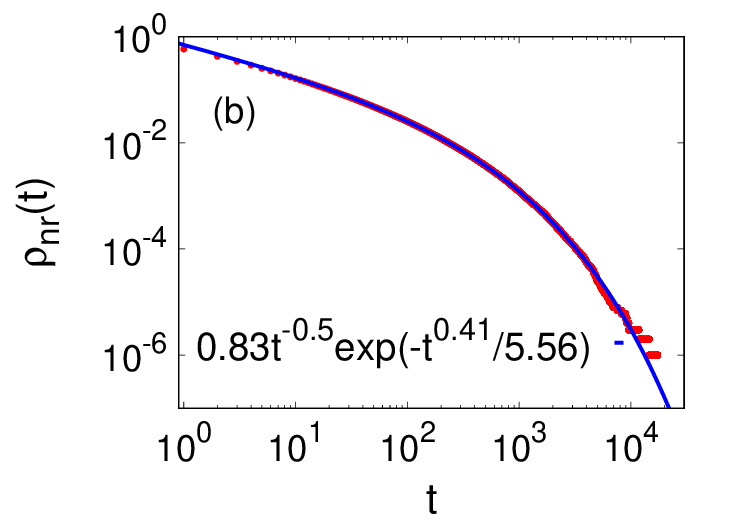}
\includegraphics[width=4cm] {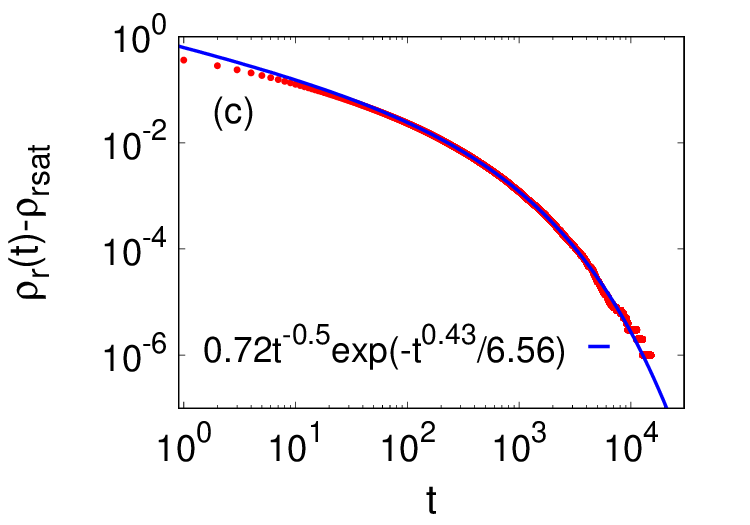}\hspace{-0.2cm}
\includegraphics[width=4cm]{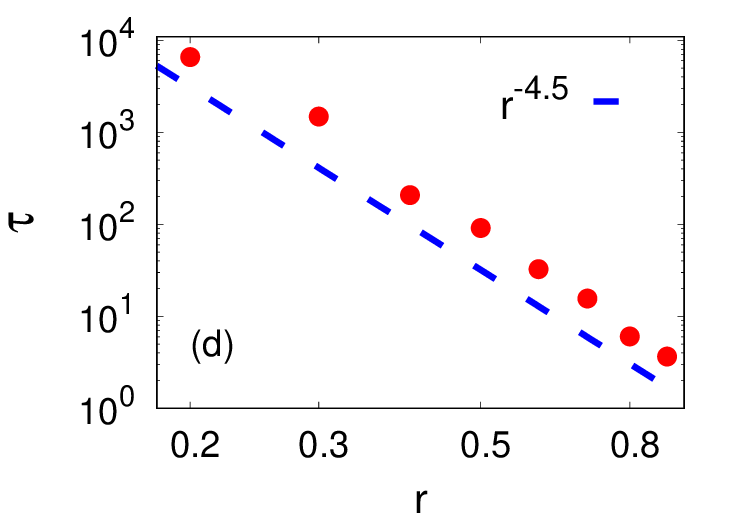}

\caption{(a) $\rho(t)-\rho_{sat}$ vs $t$. The data are fitted to Eq. \ref{rhosat_fit_generic} as mentioned in the text with the specific form mentioned in the keys. (b), (c) show $\rho_{nr}(t)$ vs $t$ and $\rho_{r}(t)-\rho_{sat}$ against $t$ respectively.  Data are fitted to the form of Eq. \ref{nonrigid_eq_ep0} and \ref{rhorsat_eq_ep0} respectively. These data are for $\epsilon=0$, $r=0.5$. (d) $\tau$ vs $r$. It shows  power law decay in $r$ with exponent nearly 4.5.} 
\label{rho_epsi0_tau}
\end{figure}

On the other hand, the
number of  rigid particles still present in the system at time $t$, scaled by their initial number, is denoted by $\rho_r(t)$ which reaches a constant value 
at the steady state. Subtracting the constant part $\rho_{rsat}$ from $\rho_r(t)$, we find this quantity follows the behaviour of Eq. \ref{rhosat_fit_generic} as
\begin{eqnarray}
\rho_{r}(t)-\rho_{rsat}=k_rt^{-0.5}\exp(-t^{b_r}/\zeta_{r}),
\label{rhorsat_eq_ep0}
\end{eqnarray}
with $k_r$ a constant and $\tau_r \propto \zeta_{r}^{1/b_r}$ the associated time scale. 
These data are shown in Fig. \ref{rho_epsi0_tau}(a),(b) and (c)
for the total, non rigid and rigid surviving particle densities respectively. Fig. \ref{rho_epsi0_tau}(d) shows the variation of the time scale $\tau$ obtained from Eq. \ref{rhosat_eq_ep0} as a function of $r$.

\begin{figure}[!ht]
\includegraphics[width=7cm]{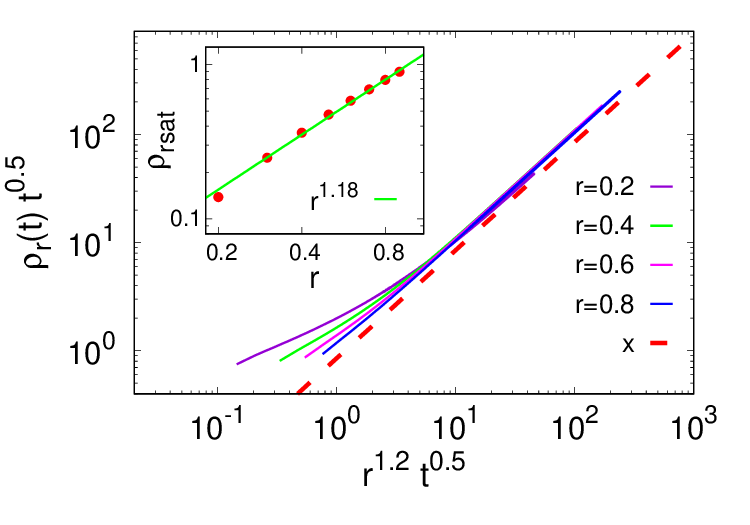}
\caption{$\rho_r(t)t^\alpha_r$ vs $r^{\beta_r} t^\alpha_r$ for $\epsilon=0$ with $\alpha_r=0.5$ and $\beta_r=1.2$. Inset shows the variation of $\rho_{rsat}$ with $r$.}
\label{rigid_sat_ep0}
\end{figure}

$\rho_{r}$ for different values of $r$ shows a  data collapse obeying Eq. \ref{rho_fit_generic} 
where $\rho_g=\rho_r$. The corresponding exponents are denoted by  $\alpha_r$ and $\beta_r$.
Fig. \ref{rigid_sat_ep0} shows that the data collapse is obtained using $\alpha_{r}=0.5$, $\beta_{r}=1.2$ and at large times, $\rho_{r}(t)t^{\alpha_{r}}$  varies linearly with $r^{\beta_{r}}t^{\alpha_r}$.
Hence, at large times, $\rho_r(t)=\rho_{rsat}$;
where  $\rho_{rsat} \sim r^{\beta_r} \sim r^{1.2}$. This power law exponent turns out to be 1.18 when $\rho_{rsat}$ is plotted against $r$ in Fig. \ref{rigid_sat_ep0}, showing consistency with the scaling form.

As mentioned earlier, here we wish to present an argument that for any $r \neq 0$, the behavior will deviate from the usual diffusive behavior. 
Let us consider an approximate theory, the independent interval approximation (IIA) \cite{naim_IIA},
using which one can obtain  results with correct scaling behavior of certain relevant quantities for  $r=0$. 
Following \cite{naim_IIA}, let $P_n(t)$ be the probability that the separation between any two domain walls is $n$, per lattice site. Hence $\sum _n n P(t)=1$ at all times as the total probability is 1. An equation governing the behavior of $P_n(t)$ can be written as 

\begin{eqnarray}
P_n(t+\Delta t)&=&[1-2\{(1-r)r+(1-r)^2\}\Delta t] \; P_n(t) \nonumber \\ 
    &+& (1-r^2) \bigg[ \Delta t \; P_{n-1}(t) \; \{1-\frac{P_1(t)}{N(t)}\}\nonumber \\
    &+&  \Delta t \; P_{n+1}(t)-\Delta t \; P_n(t) \; \frac{P_1(t)}{N(t)} \nonumber \\ 
    &+& \Delta t \; P_1(t) \sum_{i+j+1=n} \frac{P_i(t) P_j(t)}{N^2(t)}\bigg]
\end{eqnarray}

The first term on the R.H.S accounts for the probability that both domain walls do not hop. The second term contains   three contributions that occur with the coefficient $(1-r^2)$ to ensure that 
at least one domain wall is mobile. The first two of these are gain terms due to diffusion.
The last term denotes the loss due to the disappearance of the smallest domain,
located on the boundary of the domain, while the last term
represents the gain due to domain merger.
The prefactor $(1- \frac{P_1} {N})$  ensures that the hopping domain wall is not annihilated. From the above equation one gets a rate equation
\begin{eqnarray}
\frac{dP_n}{dt} &=& (P_{n-1}+P_{n+1}) \; (1-r^2) - 2 P_n \; (1-r)\nonumber \\
&+& (1-r^2) \; \frac{P_1}{N^2}\left[\sum_{i=1}^{n-2} P_i P_{n-1-i}-N(P_N+P_{n-1})\right] \nonumber 
\end{eqnarray}

Arguing that the last term is due to annihilation, the equation of motion barring that term gives
\begin{eqnarray}
\frac{dP_n}{dt}=(1-r^2)\frac{d^2P_n}{dx^2}-2P_n r (1-r)
\end{eqnarray}
which deviates from the well known diffusion equation for any $r \neq 0$.

\subsubsection{$\epsilon>0$}

\begin{figure}[!ht]
\includegraphics[width=7cm]{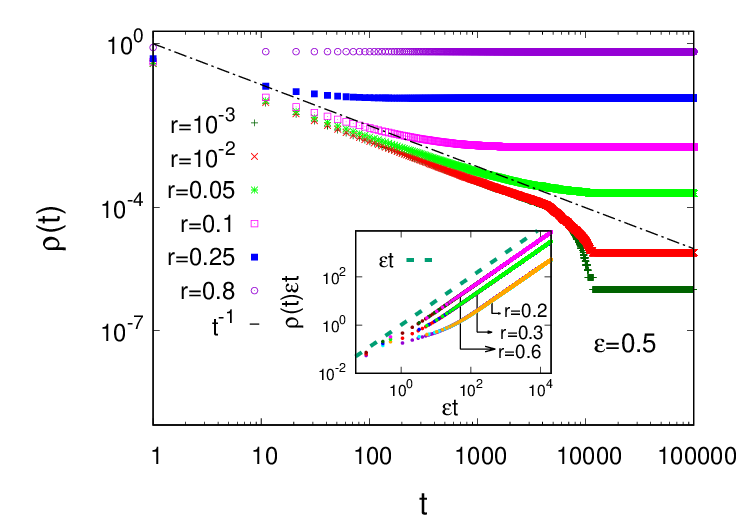}
\caption{$\rho(t)$ vs $t$ for various values of $r$ with $\epsilon=0.5$. Dashed black line indicatess $t^{-1}$.Inset shows the data collapse of $\rho(t)\epsilon t$  for $\epsilon=$ 0.1, 0.3 and 0.5 which shows a linear variation against $\epsilon t$ beyond an $r$ dependent value of $\epsi t$. Dashed green line corresponds $\epsilon t$ as mentioned in the key.}
\label{rho_ep0.5}
\end{figure}

We first present the results for the  fully deterministic case, i.e., $\epsilon=0.5$.
Fig. \ref{rho_ep0.5} shows the behavior of  $\rho(t)$  for several values of rigidity $r$. Even for very small values of $r$, we note that the value saturates. 

 Previously, it had been observed that the results are independent of $\epsi$ 
 for any value of  $\epsilon >  0 $ with a rescaling of $\epsi$ by $\epsi t$\cite{rr_pray_ps} . We note the same to be true here as 
a data collapse of the scaled variable $\rho(t)\epsi t$ can be obtained when plotted against $\epsilon t$  for a fixed value of $r$, 
shown in the inset of Fig. \ref{rho_ep0.5}. 
It is seen that beyond a value of $\epsi t$, which depends on $r$, the curves follow a linear behaviour corresponding to the saturation of $\rho (t)$
The data collapse indicates the saturation value is independent of $\epsi$ for a fixed $r$. 
Also, the saturation occurs beyond a certain value of $\epsi t $ (which is dependent on $r$), hence the time to saturate, inversely proportional to $\epsilon$, is also dependent on $r$.

For all practical  purposes we conclude that as in the case of $\epsi=0$, the results are affected for any nonzero value of $r$ although the behavior for very small $r$ values are like $r=0$ at early times (the decay occurs in a power law manner with exponent close to unity).  It will be seen later that this conclusion
is also consistent with the detailed analysis of the data.



\begin{figure}[h]
\includegraphics[width=4.5cm]{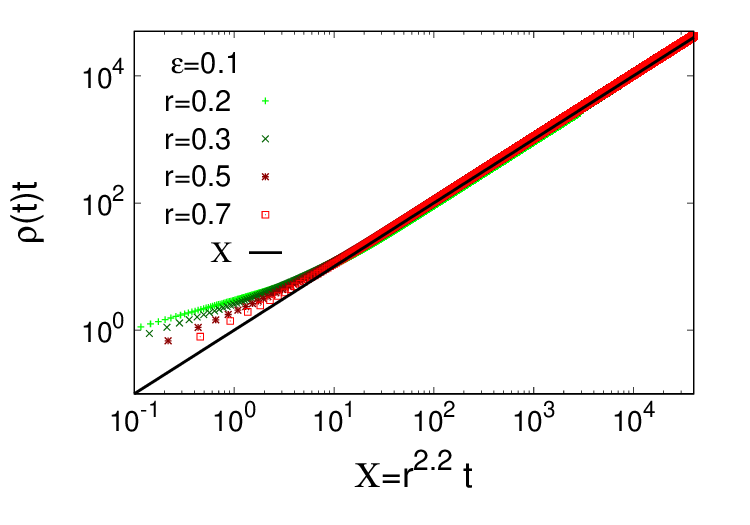} \hspace{-0.5cm}
\includegraphics[width=4.4cm]{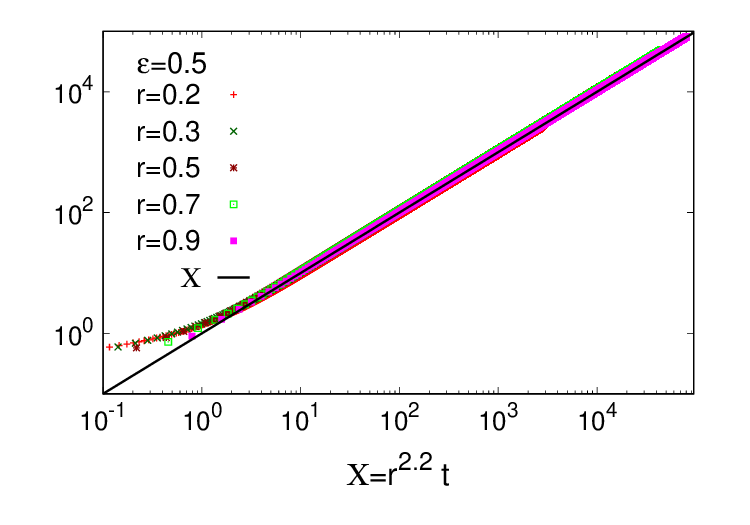}
\caption{$\rho(t)t$ vs $r^{2.2}t$ for $\epsilon=0.1 $ and $0.5$.}
\label{alive_collapse_ep0.5}
\end{figure}

For any $\epsi > 0.$, the scaled data  $\rho(t)$ can be collapsed to a single curve  for following Eq. \ref{rho_fit_generic}. However, 
the exponents have different values for $r < 0.2$ and higher values of $r$. In the higher $r$ region, the values obtained are $\alpha=1$ and $\beta=2.2$, such that
\begin{eqnarray}
\rho(t) \sim t^{-1}f(r^{2.2} t).
\label{rho_fit}
\end{eqnarray} 
Fig. \ref{alive_collapse_ep0.5} shows the collapsed data for $\epsilon=0.1$ and $0.5$ clearly 
indicating that the results do not depend on the specific value of $\epsi$ as noted above.

\begin{figure}[h]
\includegraphics[width=7cm]{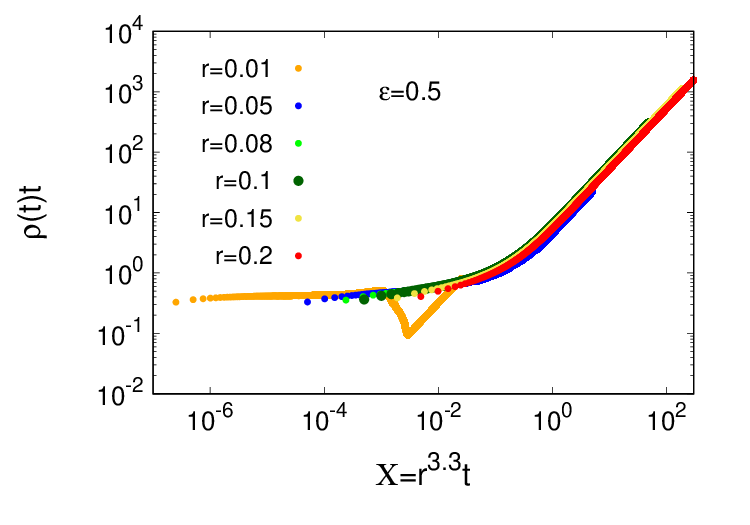}
\caption{ Data collapse of $\rho(t)t$ against $r^{3.3}t$ for   $r \leq 0.2$.}
\label{alive_collapse_smallr}
\end{figure}

A similar data collapse for smaller values of $r$ is shown in Fig. \ref{alive_collapse_smallr}, which is obtained with $\alpha =1$ and $\beta = 3.3$. The data for $r=0.01$ is an exception, and as remarked earlier, could be due to the small number of rigid walkers which are annihilated within a short time. Also, for $r=0.05$, the collapse is not appreciable due to similar reason. We note that  
the collapsed scaled values of $\rho (t)t$ data are actually  independent of the argument $r^{3.3}/t$ up to $r^{3.3}/t \sim 10^{-2}$, indicating that there is a time scale $\sim 10^{-2}/r^{3.3}$ up to which the behavior $\rho(t) \sim t^{-1}$
persists which is true for $r=0$. Hence we claim that  there is a crossover time
at low $r$ values up to which the $r=0$ behavior is followed. 
At later times, $\rho(t)$ saturates, with a value dependent on $r$ only.  
Unlike the $\epsi = 0$ case, we find that the crossover occurs at a much smaller value of $t$ so that we can observe the saturation even for small $r$ values in the simulations. 
On the other hand, the crossover to a different exponent $\beta$ occurs at 
$r \approx 0.2$. 

\begin{figure}[h]
\includegraphics[width=7cm]{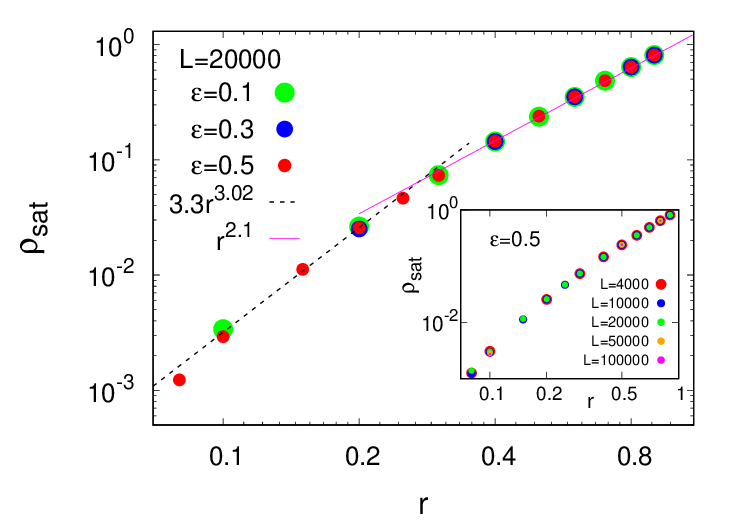}
\caption{$\rho_{sat}$ vs $r$ showing power law with two different exponents as mentioned in the key. Inset shows $\rho_{sat}$ vs $r$ for different system sizes for $\epsilon=0.5$ .}
\label{alive_sat_ep0.5}
\end{figure}

At large times $\rho(t)$ attains a constant value $\rho_{sat}$; $\rho_{sat}$ should be
proportional to  $r^{\beta \alpha}$. We find that while $\alpha=1$ for all values of $r$,
$\beta $ depends on $r$ as mentioned above.
As a result, the saturation is expected to show different behavior for small and higher ranges of $r$;   $\rho_{sat}$ should be proportional to $r^{3.3}$ and $r^{2.2}$ 
respectively for larger and smaller values of $r$.  From the numerical results we plot 
$\rho_{sat}$ against $r$ shown in Fig. \ref{alive_sat_ep0.5}. The power law behaviour indeed occurs with two different values which are close to the 
above values. We interpret this as a  crossover behavior in $r$ as the exponent changes at a value of $r \approx 0.2$.
Interestingly, $\rho_{sat}$ does not show any system size dependence as shown in the inset of Fig. \ref{alive_sat_ep0.5}, such that the crossover is not an artifact of system size dependence. 

\begin{figure}[h]
\includegraphics[width=4.5cm]{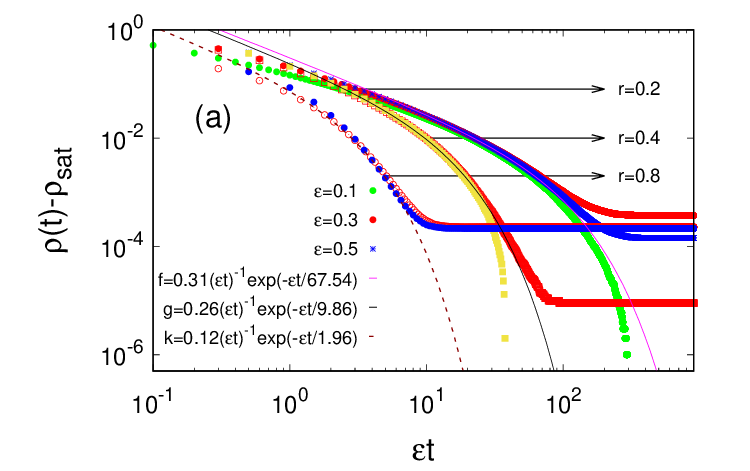} \hspace{-0.4cm}
\includegraphics[width=4cm]{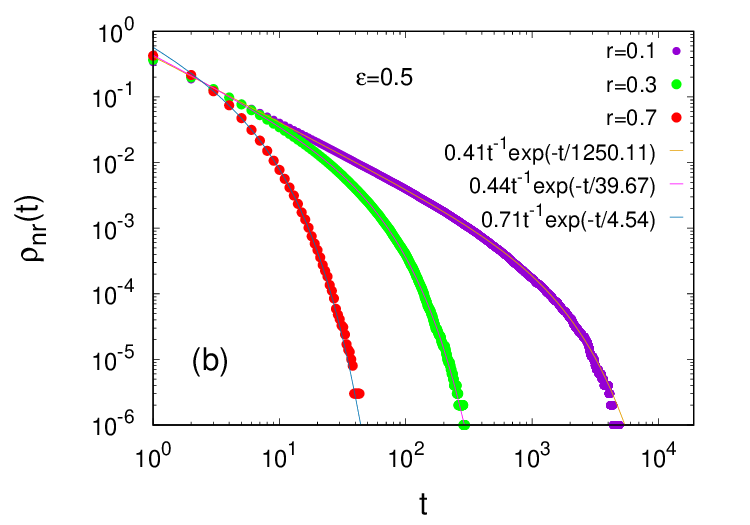}
\includegraphics[width=4cm]{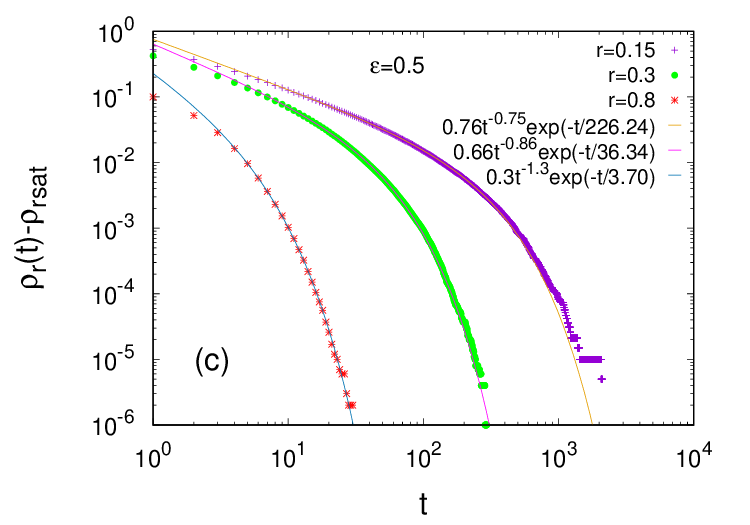} \hspace{-0.2cm}
\includegraphics[width=4cm]{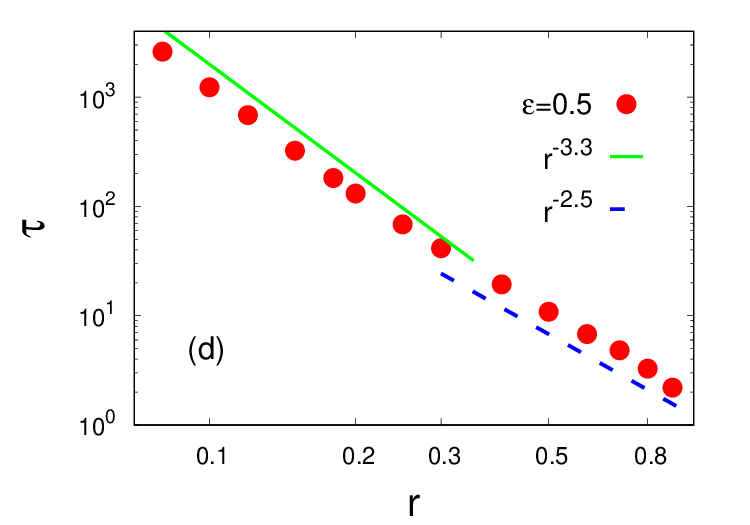}
\caption{(a) $\rho(t)-\rho_{sat}$ vs $\epsilon t$ for several $\epsilon$. The data
are fitted to Eq. \ref{rho_sat_eq} as mentioned in the key. (b), (c) show $\rho_{nr}(t)$ vs $t$ and $\rho_{r}(t)-\rho{sat}$ against $t$.  Data are fitted to the form of Eq. \ref{nonrigid_eq} and \ref{rigid_eq} respectively. These data are for
$\epsilon=0.5$. (d) $\tau$ vs $r$. It shows  power law decay in $r$ with exponent 3.3 when $r$ is small and $2.5$ for larger values of r. These data are for $L=20000$.}
\label{rho_ep0.5_tau}
\end{figure}

For $\epsilon>0$, when the saturation value $\rho_{sat}$ is subtracted from
$\rho(t)$, the data can be fit to the form of Eq. \ref{rhosat_fit_generic} with $b=1$ such that $\zeta =\tau$;
\begin{eqnarray}
\rho(t)-\rho_{sat} \sim t^{-1}\exp(-t/\tau).
\label{rho_sat_eq}
\end{eqnarray}
$\tau$ can be identified as a timescale in the system. A data collapse for all $\epsilon$ can be obtained by scaling the time $t$ with $\epsilon$, such as
\begin{eqnarray}
\rho(t)-\rho_{sat}=k^\prime(\epsilon t)^{-1}\exp(-\epsilon t/\tau_c).
\label{rho_sat_epsieq}
\end{eqnarray}
The data are shown in Fig. \ref{rho_ep0.5_tau}. Here also, we find that $\tau$ decreases in a power law manner with $r$ with the two different exponent values for $r < 0.2$ and $r> 0.2$ and the numerical values are close to the theoretically predicted values $\beta/\alpha$, as shown in Fig. \ref{rho_ep0.5_tau}(d).

The decay of $\rho_{nr}(t)$ for $\epsilon>0$ follows the behaviour of Eq. \ref{rhosat_fit_generic} with $a=1$ and  zero saturation value;

\begin{eqnarray}
\rho_{nr}(t)=k_{nr}^\prime  t^{-1}\exp(-t/\tau_{nr}).
\label{nonrigid_eq}
\end{eqnarray}



In this case also, $\rho_r(t)$, reaches a constant value after
a certain time. In the similar way as for $\epsilon=0$, subtracting the constant part $\rho_{rsat}$, we have an equation similar to Eq. \ref{rhosat_fit_generic}: 
\begin{eqnarray}
\rho_{r}(t)-\rho_{rsat}=k_r^\prime t^{-1}\exp(-t/\tau_{r}).
\label{rigid_eq}
\end{eqnarray}



\begin{figure}[h]
\includegraphics[width=7cm]{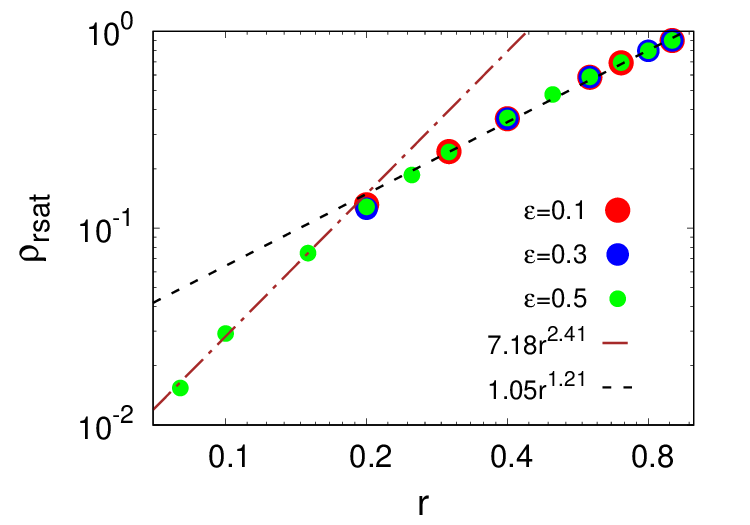}
\includegraphics[width=7cm]{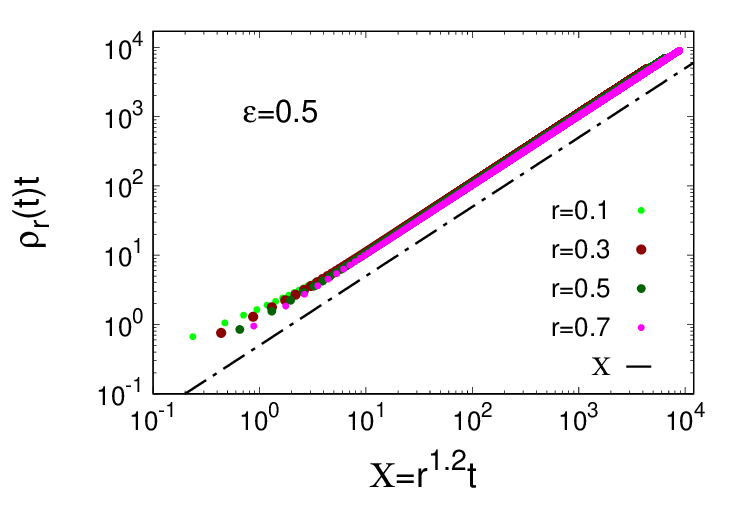}
\caption{$\rho_{rsat}$ vs $r$ (upper panel) shows a power law variation. Lower panel: Data collapse for the rigid walls: $\rho_{r}(t)t$ vs $r^{1.2}t$.}
\label{rigid_sat}
\end{figure}

The saturation value of $\rho_{r}$ shows a power law growth with $r$ with an exponent $\sim 1.21$ for larger values of $r$ ($r \geq 0.2$). Fig. \ref{rigid_sat}
shows the results.  $\rho_{r}(t)t$ shows a similar data collapse as Eq. \ref{rho_fit} against $r^{\beta_r}t$
 such that $\rho_{r}(t)t \propto t^{-\alpha_{r}}f^\prime(r^{\beta_{r}} t)$. Fig. \ref{rigid_sat} (lower panel) shows $\alpha_{r}=1$ and $\beta_{r}=1.2$. Hence
$\rho_{rsat}$ will show a similar behaviour, $\rho_{rsat} \sim r^{\beta_{r} \alpha_{r}} \sim r^{1.2}$. The power law exponent turns out to be 1.21 as shown in Fig. \ref{rigid_sat}, upper panel.

Before ending this section, let us add an argument to show how $\beta$ and $\beta_r$ are related:\\
As per our definitions,
\begin{equation}
 \rho (t) = r \rho_r(t) + (1-r) \rho_{nr} (t).  
\end{equation}
Hence, since the non-rigid walkers vanish at large times, 
the saturation values of $\rho(t)$
coincides with that of $\rho_{r}(t)$ ($\rho_{sat}=r\rho_{rsat}$) leading to the relation
\begin{equation}
 \beta_r =  \beta -1.   
\end{equation}
Indeed, the results obtained for both $\epsi = 0$ and $\epsi > 0$ 
are consistent with the above equation.



\subsection{Persistence $P(t)$}
We next present the results for the persistence probability $P(t)$ defined earlier in section \ref{model}.
In the absence of rigid particles,  $P(t)$ shows a power law behaviour in time, $P(t) \sim t^{-\theta}$
where $\theta=0.375$ for $\epsilon=0$ (an exact result) and $\theta\simeq 0.235$ for $\epsilon>0$ (obtained numerically). 
We observe that in general the scaled persistence probability $P(t)t^\theta$ shows a data collapse when $P(t)t^\theta$ is plotted against $r^\gamma t$. Therefore, the scaling form can be written in general as

\begin{eqnarray}
    P(t)\sim t^{-\theta}f(r^\gamma t^\theta)
    \label{persis_eq}
\end{eqnarray}
We denote the scaling argument by $X=r^\gamma t^\theta$. The results for $\epsi=0$ and $\epsi > 0$ are reported in the two following subsections.

\subsubsection{$\epsilon=0$}

\begin{figure}[h]
\includegraphics[width=6cm]{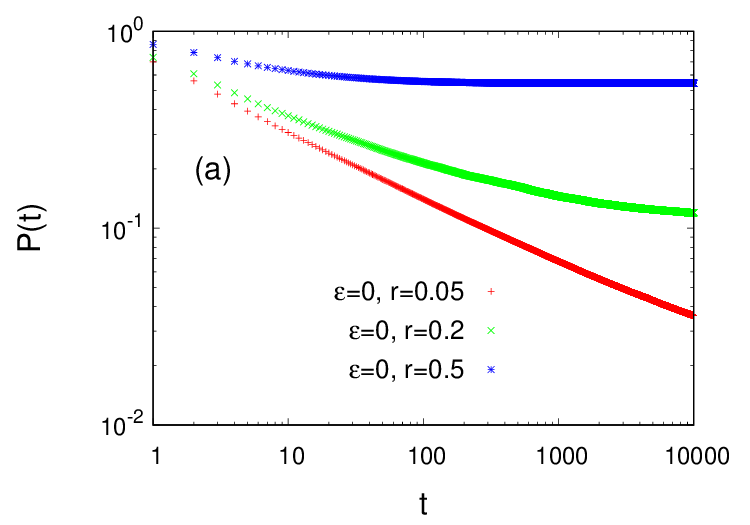}
\includegraphics[width=6cm]{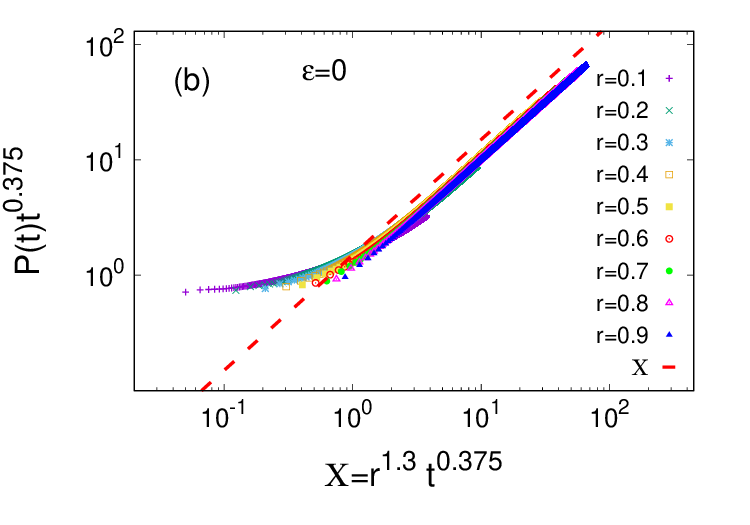}
\caption{(a)$P(t)$ against $t$ for $\epsilon=0$. (b) $P(t)t^{0.375}$ against $r^{1.3}t^{0.375}$.}
\label{persis_fig_ep0}
\end{figure}

The persistence probability for $\epsilon=0$ is shown in Fig. \ref{persis_fig_ep0} for several values of $r$.  The saturation value is reached rapidly when $r$ is large.

 A data collapse can be obtained as mentioned in Eq. \ref{persis_eq} with $\theta \simeq 0.375$ and $\gamma \simeq 1.3$. We note the scaling function $f(X)$,  up to $X \approx 1$,  is fairly constant indicating the behavior $P(t) \propto t^{-0.375}$ as in the $r=0$ case. The scaling function is linear for larger values of the scaling argument which implies $P(t)$ shows a saturation behavior with its magnitude increasing as  $r^{1/3}$. Thus we have a crossover behavior occurring at $X\approx 1$.


\subsubsection{$\epsilon>0$}

\begin{figure}[h]
\includegraphics[width=6cm]{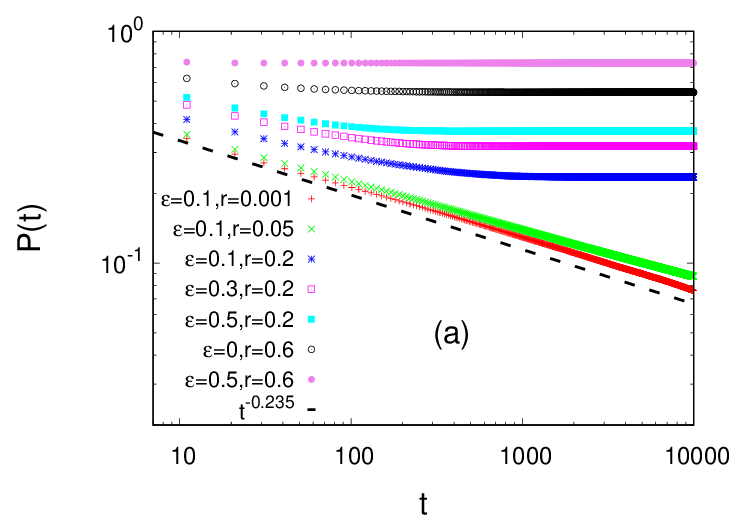}
\includegraphics[width=6cm]{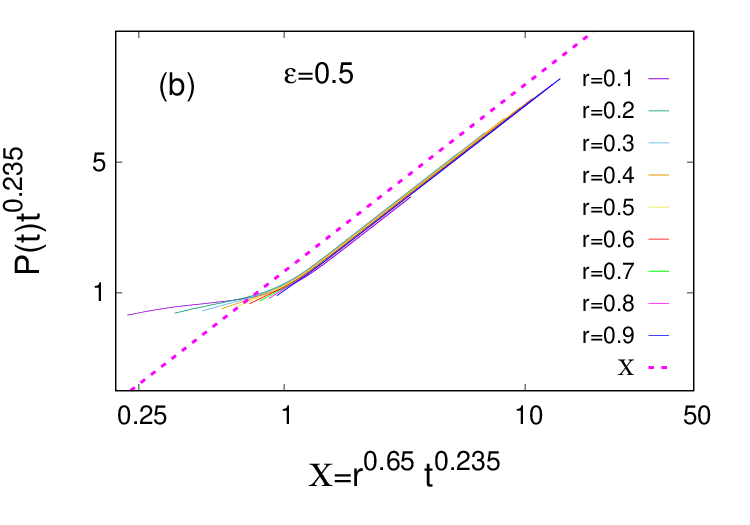}
\caption{(a) $P(t)$ against $t$ for $\epsilon=0.5$. (b) $P(t)t^{0.235}$ against $r^{0.65}t^{0.375}$.}
\label{persis_fig_ep0.5}
\end{figure}

For $\epsilon>0$, $P(t)$ shows a power law decay with an exponent which agrees fairly well with the $r=0$ value,  0.235, when $r$ is very small. At later times, it shows a saturation behavior implying the qualitative behavior of $P(t)$ is independent of $\epsi$. For larger values of $r$, the saturation is reached very fast for $\epsilon>0$ compared to the $\epsilon=0$ case. Fig. \ref{persis_fig_ep0.5}(a) shows the plot.

A similar data collapse can also be obtained as Eq. \ref{persis_eq} for $\epsilon>0$. Fig. \ref{persis_fig_ep0.5} shows the data. The data collapse suggests that the value of $\theta$ and $\gamma$ are close to 0.235 and 0.65. 
 Similar to the $\epsi=0$ case, we find a crossover, occurring at $X=r^{0.65}t^{0.235}=1$ for $\epsi = 0.5$ in particular. 
Fig. \ref{persis_fig_ep0.5}(b) shows the results.

Interestingly, the exponent $\gamma$ is very close to half of that for $\epsilon = 0$. Since $r$ is in the range $(0,1)$, the increase of the saturation value of $P(t)$ with $r$ is faster for $\epsi > 0$. The curvature is also different in the two cases. 
We have checked that a data collapse for all values of $\epsilon$ can be obtained when $P(t)t^\theta$ is plotted against $\epsilon t$.




\begin{table}[h]
 
    \centering
    \begin{tabular}{|c|c|c|}
        \hline
        & $\epsilon=0$ & $\epsilon \neq 0$ \\
        \hline

$\alpha$, $\beta$ &  0.5, 2.2 ($r > 0.2)$ & 1.0, 2.2 ($r > 0.2$)\\
&   & 1.0, 3.3 ($r < 0.2$)\\
        \hline
       $\alpha_r$, $\beta_r$ &  0.5, 1.2 ($r > 0.2$)  & 1, 1.2 ($r > 0.2$)\\
         &  & 1, 2.2  ($r < 0.2$) \\
        \hline

        $\theta$ & 0.375 & 0.235  \\
        \hline
        $\gamma$ &  1.3 & 0.65  \\
        
        \hline
       
        \hline
    \end{tabular}
    \caption{Exponent values obtained from the present study}
    \label{Table1}
\end{table}

\section{summary and conclusions}

In a  model in which particles perform a walk determined by a dynamic bias and are annihilated 
when they meet, rigid particles are introduced which are immobile but can be annihilated. The model is a toy model to mimic the depinning transition, if any, in a material with impurities. The non-rigid particles  move towards their nearest neighbour with a finite probability $0.5+\epsilon$, such that for $\epsilon \neq 0$, there is a dynamic bias. 

With a fraction $r$ of the total number of particles occupying the sites of a one dimensional chain, we find that any $r \neq 0$ affects the scaling behavior of the surviving particle densities with time. While the dominant behavior is still a power law of the form $t^{-\alpha}$,
where $\alpha$ is the exponent when $r=0$, a scaling function with the argument $r^\beta t^\alpha$ modulates the decay behavior. Interestingly, $\beta$ has two different values for $r > 0.2$ and $r < 0.2$ as clearly found for $\epsi \neq 0$ while for the $\epsi =0$, one can only explore the larger $r$ region. At least in this region, though $\alpha$ values are different for $\epsi =0$ and $\epsi>0$ (already known for  $r=0$), the $\beta$ values are same for any $\epsilon \geq 0$. Along with the total number of particles, one can also separately study the time dependence of the 
densities of the rigid and nonrigid particles. The exponent $\beta_r$ for the rigid particles is  $\beta-1$ as argued theoretically and also obtained numerically.

An interesting crossover behavior in time is also found for both $\epsi =0$ and $\epsi \neq 0$ to show that the pure power law behavior crosses over to a saturation behavior for $r \neq 0$. However, for $\epsi =0$, the crossover time for small $r$ values turn out to be too large to study by numerical simulations and that is why the exponent values are only reported for larger $r$ values. 

We also estimate the relevant time scales for the different densities. The timescale obtained from the behaviour of the total densities of the particle shows a power law behavior with $r$ where the exponent is related to $\alpha$ and $\beta$.

Another feature, namely, the persistence behavior is also studied. Again, the power law behavior is found to be modulated by a scaling function which involves another exponent, effective only when $r \neq 0$. Interestingly, the exponent for $\epsi =0$ is exactly twice of that for $\epsi \neq 0$. Apparently, this exponent $\gamma$ is unrelated to $\beta$. All the exponent values obtained have been summarised in Table \ref{Table1}.

The important finding is, at large times, even for small $r$, a finite number of rigid particles remain in the system. This happens as both the rigid and non-rigid particles are annihilated when they meet. The result is reminiscent of the vanishing of the order parameter at any finite
temperature in one dimensional Ising model. In fact, if we think of the rigid particles to be  separating domains of different orientation, their presence would ensure absence of order with a high probability.

Finally we add that if rigidity is introduced in the system in an annealed matter such that the at every time step different walkers become rigid, $\rho(t) \propto t^{-\alpha}$ behavior is recovered for any $r<1$; however, the prefactor increases with $r$.\\

Acknowledgment: RR acknowledges IMSc Post doctoral Fellowship, The numerical computations were carried out on the Nandadevi cluster, maintained and supported by the Institute of Mathematical Science’s High-Performance Computing Center. PS acknowledges financial support from CSIR. \\

\end{document}